\documentclass[prc,twocolumn,groupaddress]{revtex4}

\usepackage{epsfig}
\usepackage{graphicx}
\usepackage{dcolumn}
\usepackage{bm} 
\usepackage{epstopdf}
\usepackage{amsmath}
\usepackage{amsfonts}
\usepackage[colorlinks,linkcolor=blue,anchorcolor=blue,citecolor=blue]{hyperref}
\newcommand{\beq}{\begin{equation}}
\newcommand{\eeq}{\end{equation}}
\newcommand{\beqa}{\begin{eqnarray}}
\newcommand{\eeqa}{\end{eqnarray}}

\def\jpb#1{{ J.\ Phys.\ B} {\bf#1}}

\def\jpg#1{{ J.\ Phys.\ G} {\bf#1}}

\def\natphys#1{{ Nature\ Phys.\ } {\bf#1}}
\def\natcomm#1{{ Nat.\ Commun.} {\bf#1}}

\def\pra#1{{ Phys.\ Rev. A\/} {\bf#1}}

\def\prc#1{{ Phys.\ Rev. C\/} {\bf#1}}

\def\pre#1{{ Phys.\ Rev. E\/} {\bf#1}}
\def\prl#1{{ Phys.\ Rev.\ Lett.} {\bf#1}}
\def\prx#1{{ Phys.\ Rev. X\/} {\bf#1}}
\def\rmp#1{{ Rev.\ Mod.\ Phys.} {\bf#1}}

\def\nat#1{{ Nature} {\bf#1}}

\begin{document}

\title{Substantially enhanced deuteron-triton fusion probabilities in intense low-frequency laser fields}

\author{Xu Wang}
\email{xwang@gscaep.ac.cn}
\affiliation{Graduate School, China Academy of Engineering Physics, Beijing 100193, China}

\date{\today}

\begin{abstract}

Deuteron-triton (DT) fusion is the primary fusion reaction used in controlled fusion research, mainly for its relatively high reaction cross sections compared to other fusion options. Even so, to attain appreciable reaction probabilities very high temperatures (on the order of 10-100 million kelvins) are required, which are extremely challenging to achieve and maintain. We show that intense low-frequency laser fields, such as those in the near-infrared regime for the majority of intense laser facilities around the world, are highly effective in transferring energy to the DT system and enhancing the DT fusion probabilities. The fusion probabilities are shown to be enhanced by at least an order of magnitude in 800-nm laser fields with intensities on the order of 10$^{21}$ W/cm$^2$. The demanding temperature requirement of controlled nuclear fusion may be relaxed if intense low-frequency lasers are exploited.

\end{abstract}

\maketitle

{\it Introduction.} Controlled nuclear fusion has the potential of supplying sustainable and clean energy solutions. In either magnetic confinement fusion \cite{Hinton-76, Sheffield-94, Ongena-16} or inertial confinement fusion \cite{Lindl-04, Hurricane-14, Betti-16}, which are two major schemes for controlled fusion research, the DT fusion reaction (D + T $\rightarrow$ $^4$He + n + 17.6 MeV) is chosen for its relatively high reaction cross sections compared to other fusion options \cite{Belyaev-05, Labaune-13, Picciotto-14}. Even so, the required temperature is very high, usually on the order of 10-100 million kelvins, in order to attain appreciable fusion reaction probabilities. These temperatures are very challenging to achieve and maintain. New methods or tools that can further increase the DT reaction probabilities and relax the demanding temperature requirement are therefore particularly desirable.

Intense laser is a potential candidate. It is one of the few tools that we possess to generate extreme conditions. Rapid progresses have been achieved on intense laser technologies since the invention of the chirped pulse amplification technique \cite{CPA}. Light with intensities on the order of 10$^{21}$ to 10$^{22}$ W/cm$^2$ can be generated nowadays, and further increase for another one or two orders of magnitude can be expected in the near future, for example, with the Extreme Light Infrastructure of Europe \cite{ELI-1, ELI-2}. The possibility of using intense laser fields to influence nuclear processes, such as alpha decay, is intriguing and has attracted attention \cite{Cortes-2013, Misicu-2013, Delion-2017, Qi-19, Palffy-19}.

Whether intense laser fields can enhance the DT fusion probabilities remains unclear, yet this question is not only important for controlled fusion research but also intriguing on its own. In the limit of very high laser frequencies as in the X-ray regime, qualitative estimations have been given by Queisser and Sch\"utzhold using a Floquet scattering method \cite{Queisser-19} and by Lv et al. using a Kramers-Henneberger approximation \cite{Lv-19}, indicating positive answers. However, these methods or approximations cannot be applied to lasers in the near-infrared regime for the majority of intense laser facilities. The difficulty originates from the large number of photons involved (e.g. exceeding 10,000) when high intensity combines with low frequency (photon energy). In the X-ray regime, by contrast, the involved number of photons is very limited, permitting simplifications.

The goal of the current paper is to answer the questions whether, how, and by how much the DT fusion probabilities can be enhanced by intense laser fields. We present a physically intuitive analysis that is capable of including the large numbers of involved photons. The results show that intense low-frequency laser fields are highly effective in transferring energy to the DT system and enhancing the fusion probabilities. This effectiveness is attributed to the energy properties of the Volkov state, the quantum state of a charge particle in an electromagnetic field. The DT fusion probabilities are shown to be enhanced by at least an order of magnitude in 800-nm laser fields with intensities on the order of 10$^{21}$ W/cm$^2$. The results also show that low-frequency lasers are more efficient in enhancing DT fusion than high-frequency lasers. The demanding temperature requirement of controlled fusion may be relaxed if intense low-frequency laser fields are exploited.

\begin{figure} [t!]
 \centering
 \includegraphics[width=4.2cm,trim=0 0 0 0]{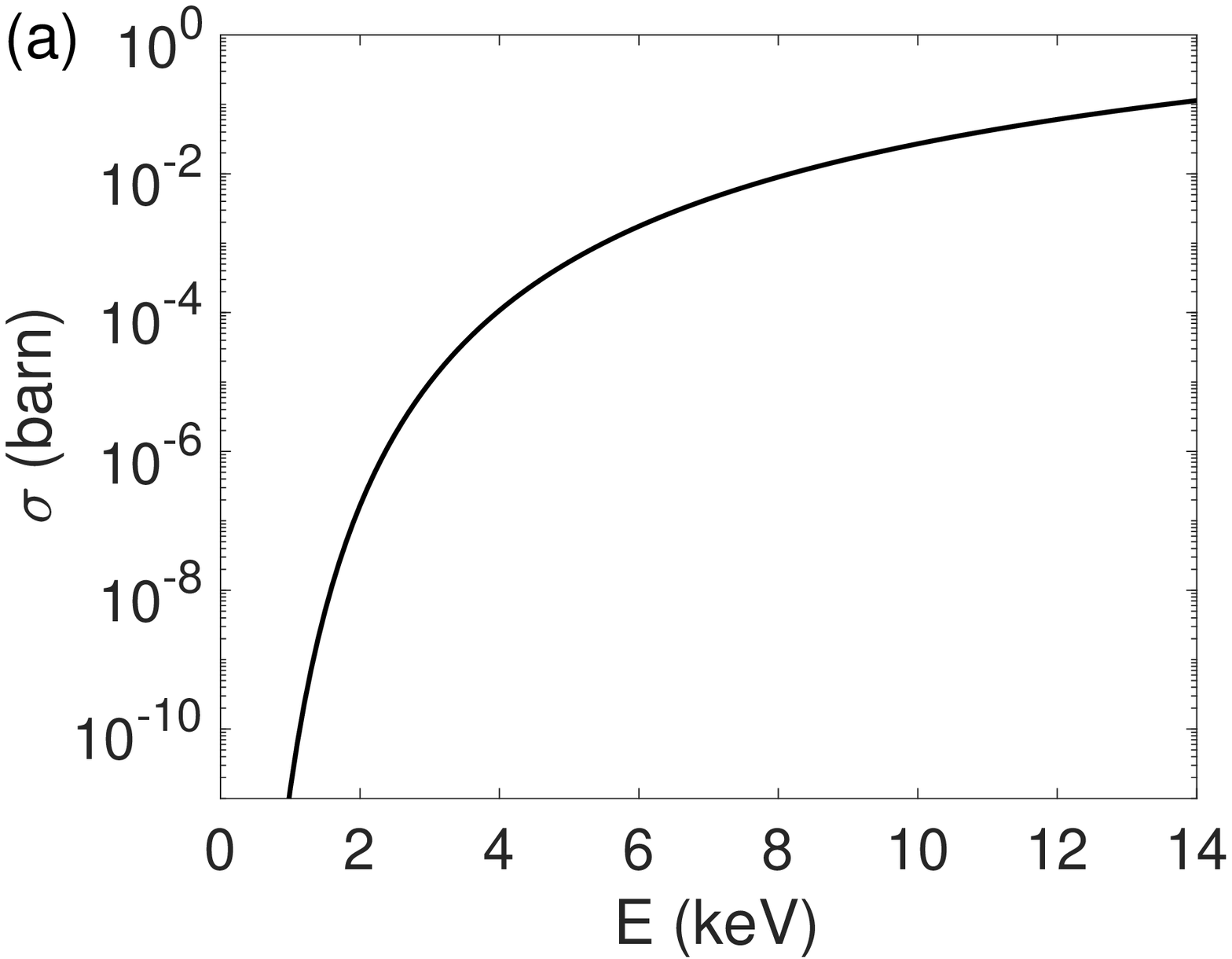}
 \includegraphics[width=4.2cm,trim=0 0 0 0]{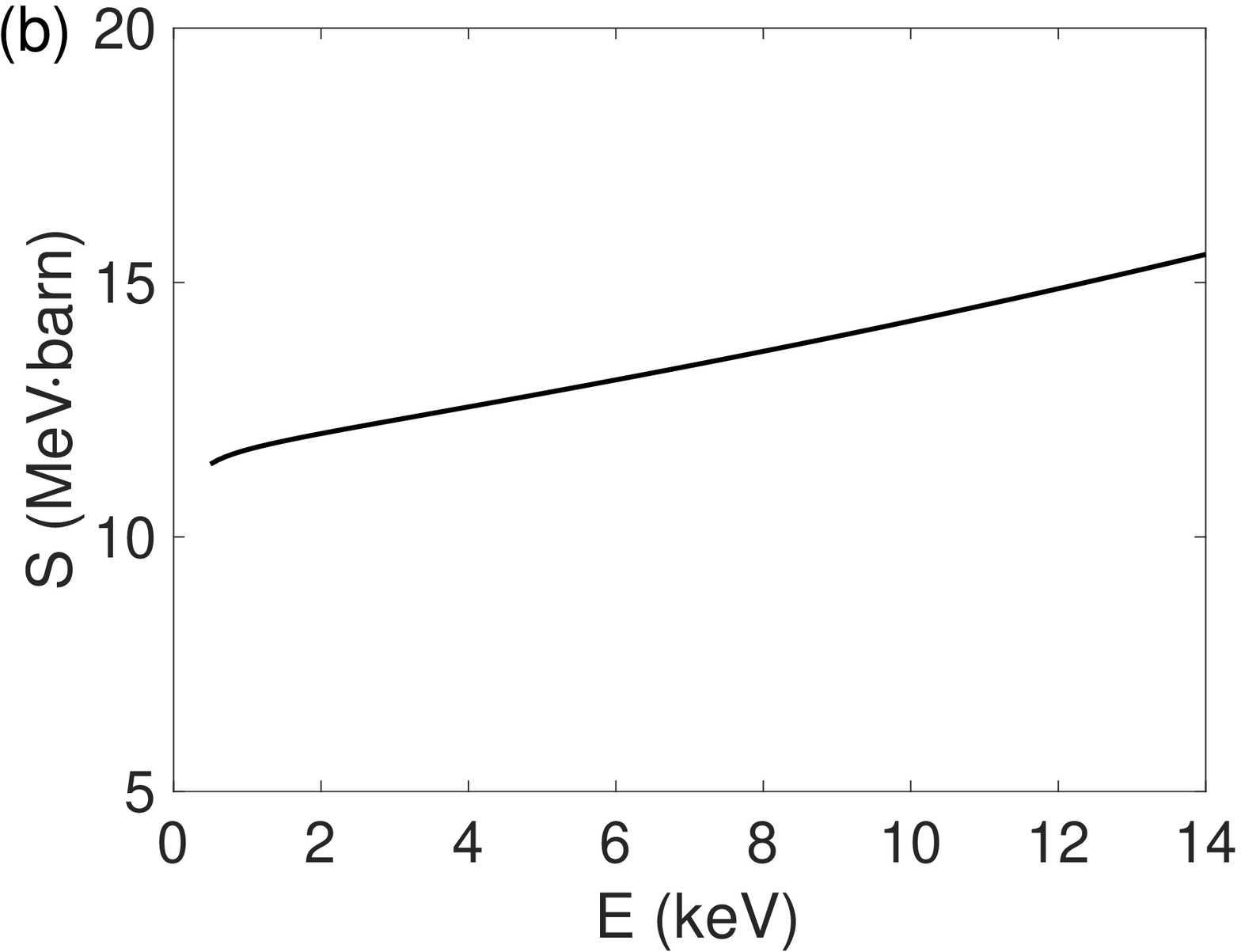}
 \caption{(a) Cross section of DT fusion as a function of relative-motion energy $E$. (b) The corresponding $S$ function.}\label{f.sigmaSE}
\end{figure}

{\it DT fusion without laser fields.} We first give a brief description of the DT fusion cross section without the presence of laser fields. The cross section can be written in the following three-factor form \cite{Burbidge-57}
\beq
\sigma(E) = S(E) \frac{1}{E} \exp\left(-\frac{B_G}{\sqrt{E}}\right), \label{e.CS}
\eeq
where $E$ is the energy in the center-of-mass frame. The term $1/E$ is called the geometrical factor. The exponential term is the probability of tunneling \cite{Gamow-28} through the repulsive DT Coulomb barrier, and $B_G = 34.38 \sqrt{\text{keV}}$ is called the Gamow constant. The slowly-varying function $S(E)$ describes the nuclear physics when deuteron and triton are very close and nuclear potentials are in effect. Here we use the parametrization given by Bosch and Hale \cite{Bosch-Hale} 
\beq
S(E) = \frac{ A_1 + E(A_2+E(A_3+EA_4)) }{ 1+E(B_1+E(B_2+E(B_3+EB_4))) }, \label{e.SE}
\eeq
which yields accurate agreements to experimental data, especially for relatively low energies that are of relevance to controlled fusion research. The values of the parameters $A_i$'s and $B_i$'s can be found in Table IV of \cite{Bosch-Hale} and will not be repeated here. The cross section and the $S$ function are plotted in Fig. \ref{f.sigmaSE} for energies below 14 keV.

\begin{figure*} [t!]
 \centering
 \includegraphics[width=5cm,trim=0 0 0 0]{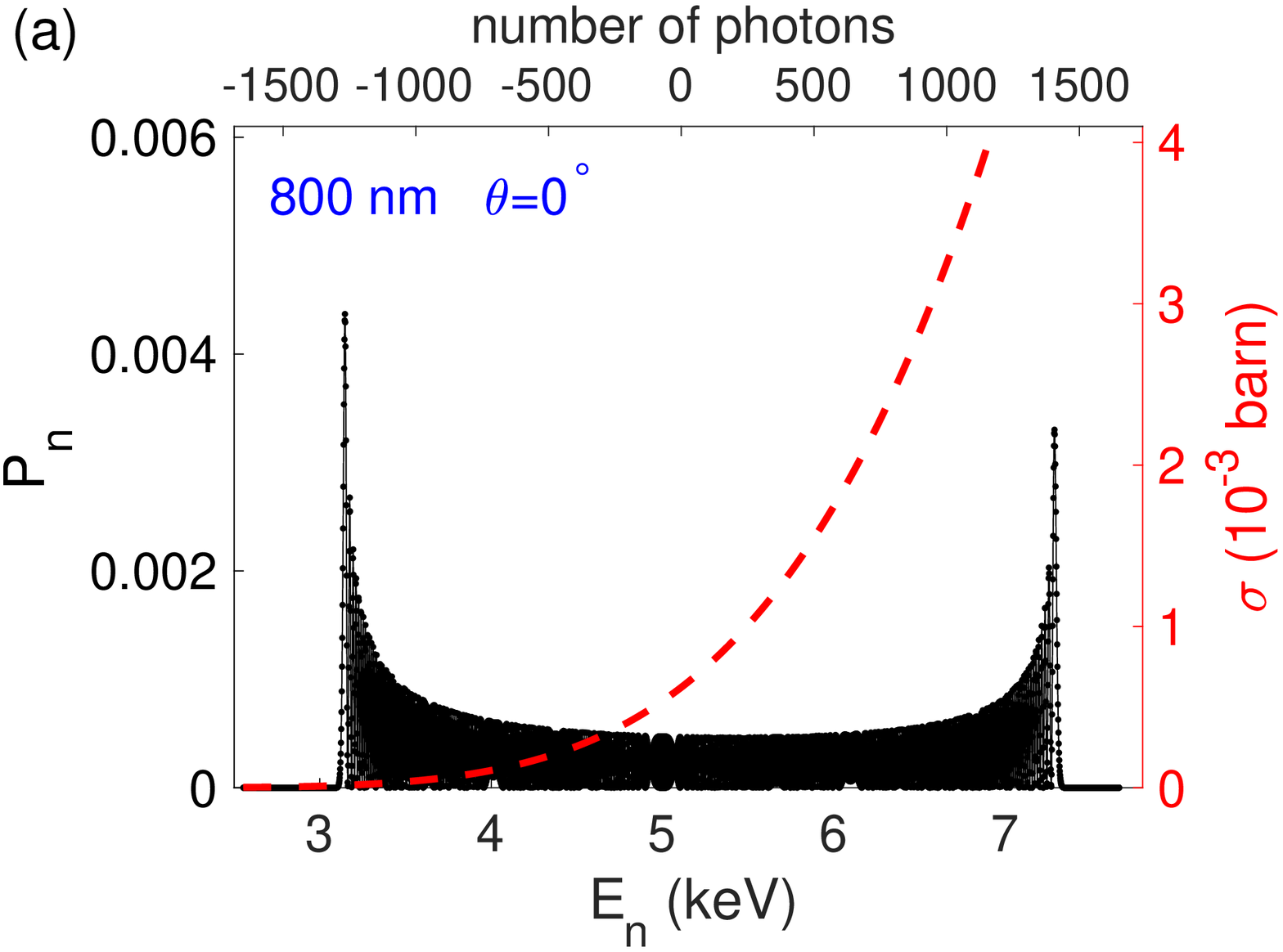}
 \hspace{0.1cm}
 \includegraphics[width=5cm,trim=0 0 0 0]{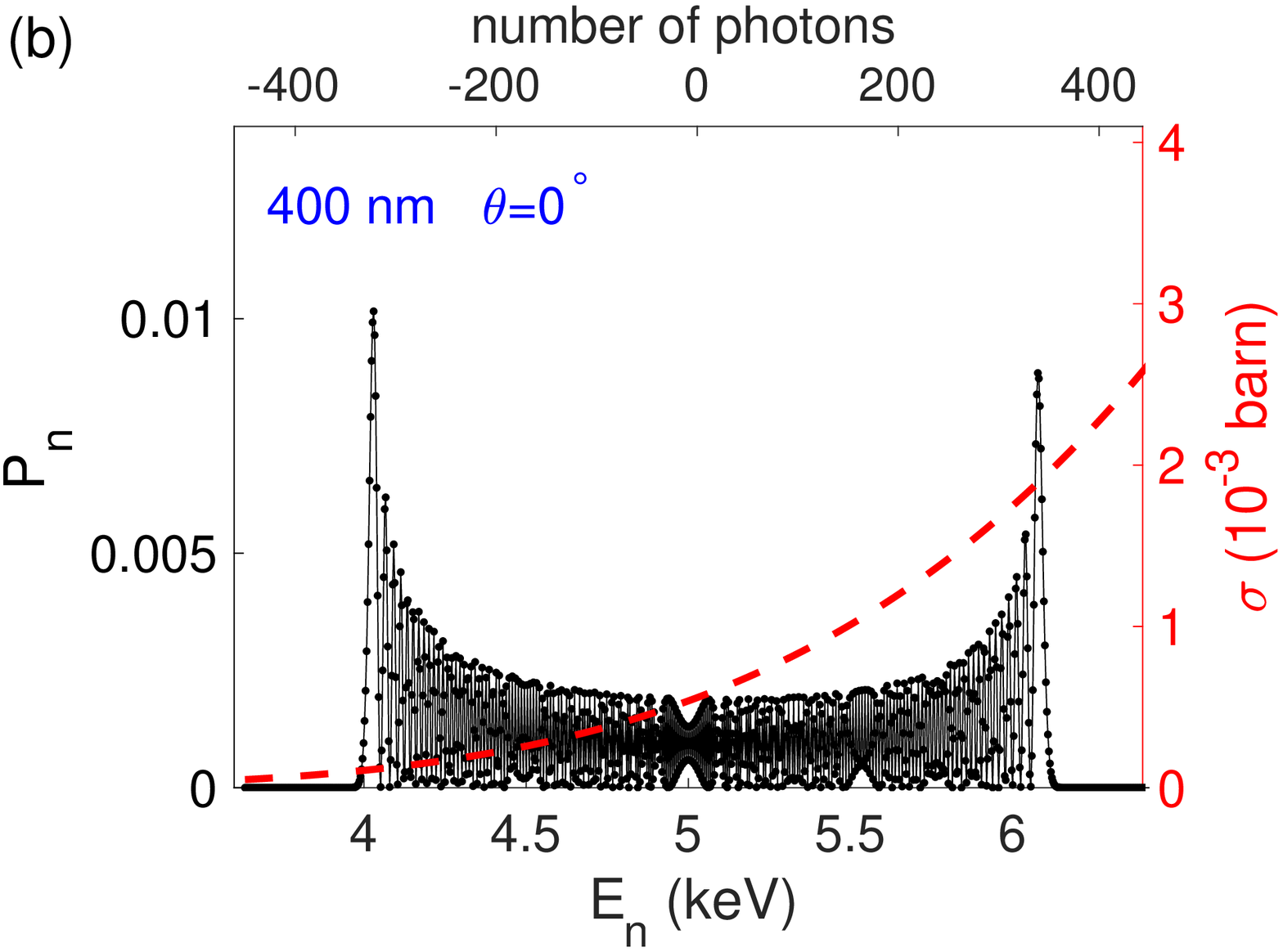}
 \hspace{0.1cm}
 \includegraphics[width=5cm,trim=0 0 0 0]{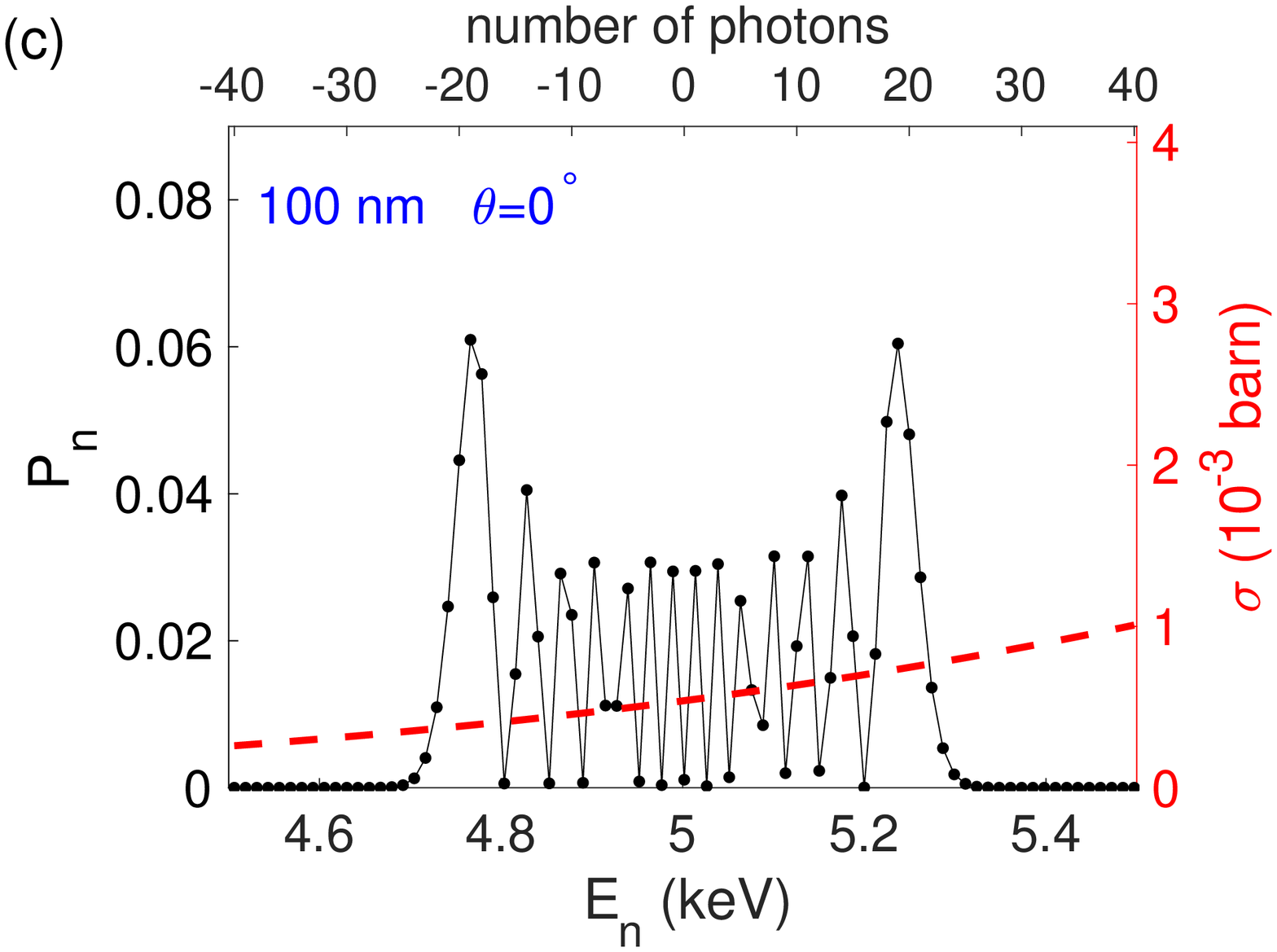} \\
 \vspace{0.1cm}
  \includegraphics[width=5cm,trim=0 0 0 0]{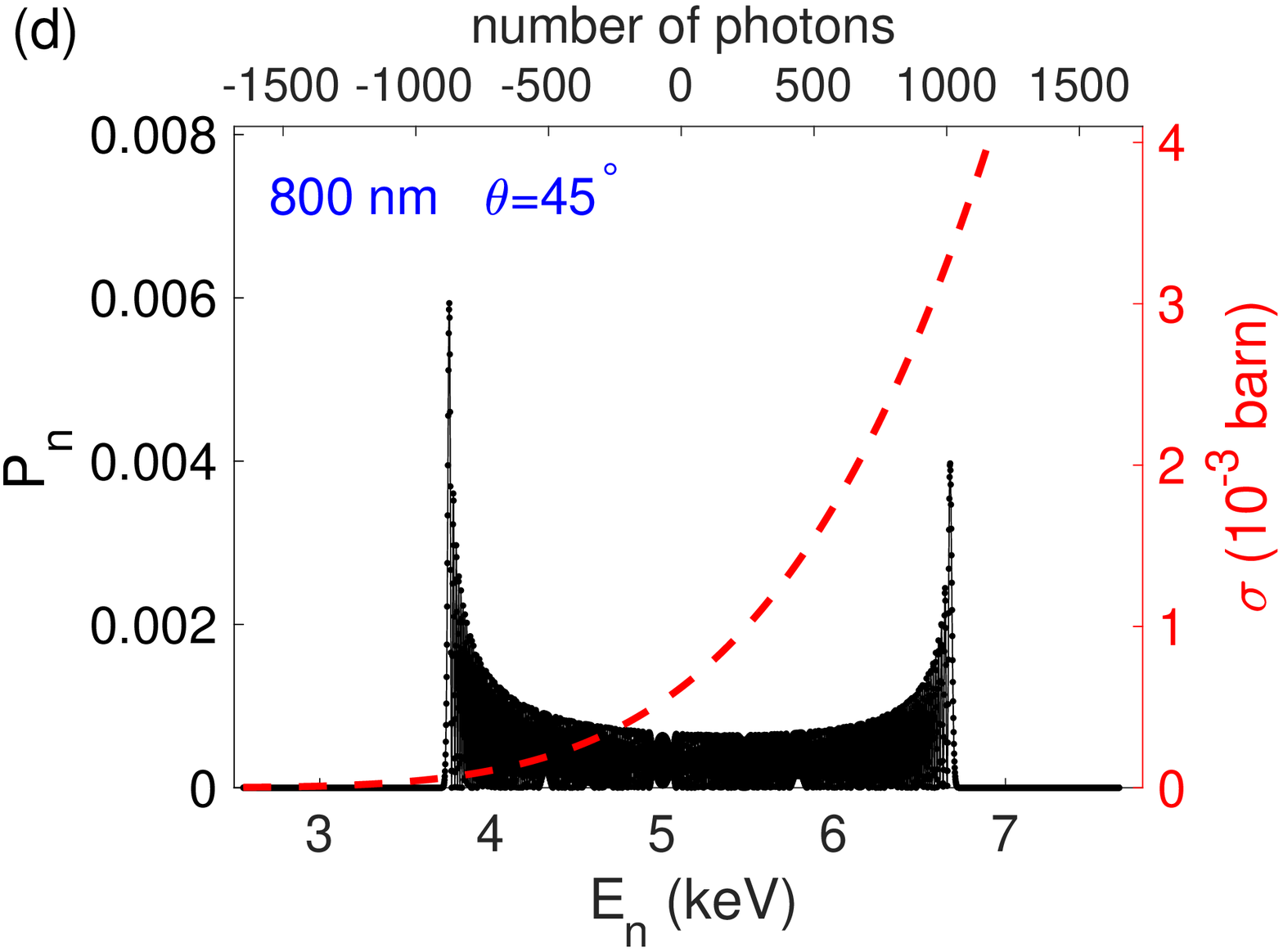}
  \hspace{0.1cm}
 \includegraphics[width=5cm,trim=0 0 0 0]{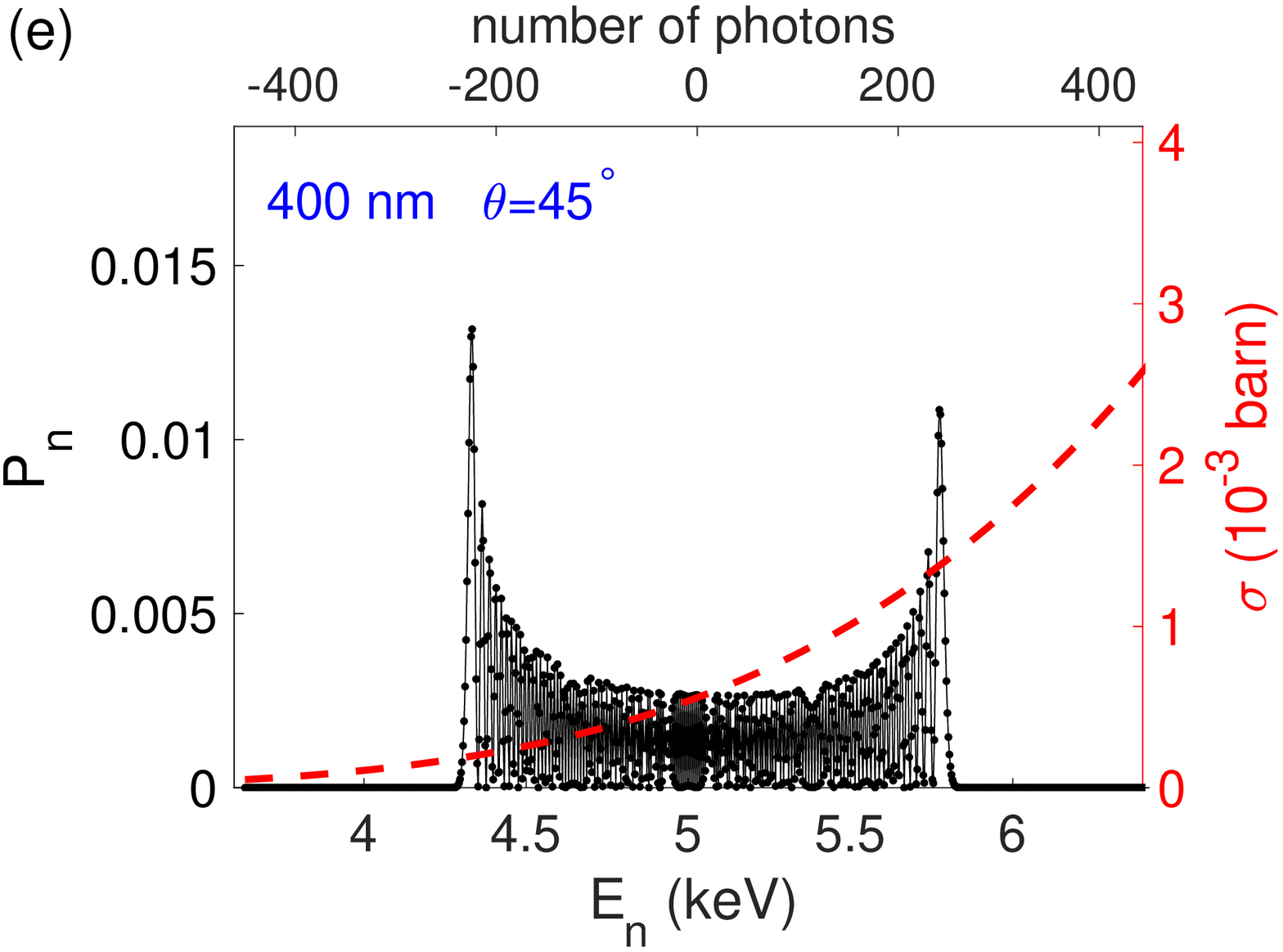}
 \hspace{0.1cm}
 \includegraphics[width=5cm,trim=0 0 0 0]{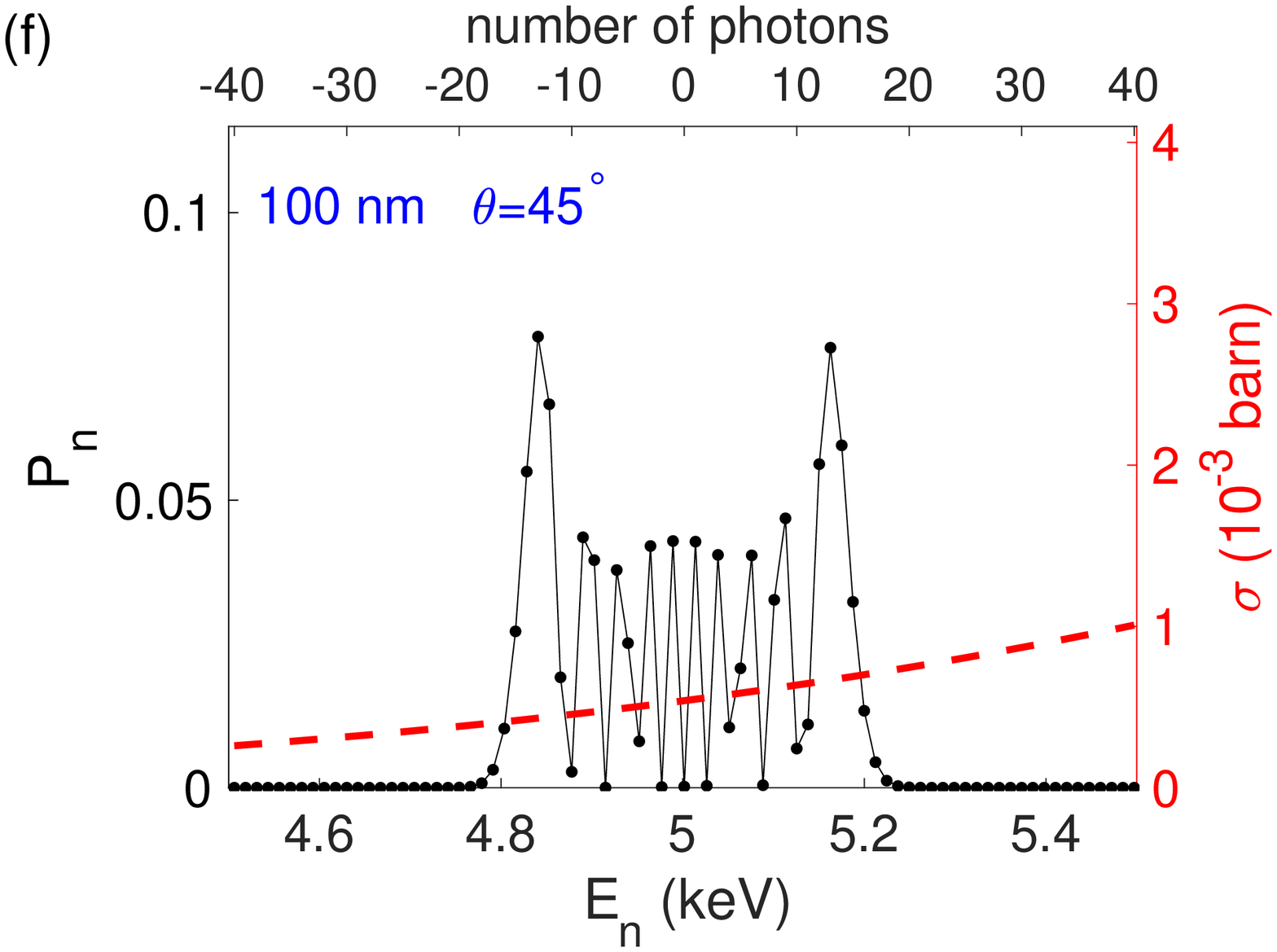}\\
 \vspace{0.1cm}
  \includegraphics[width=5cm,trim=0 0 0 0]{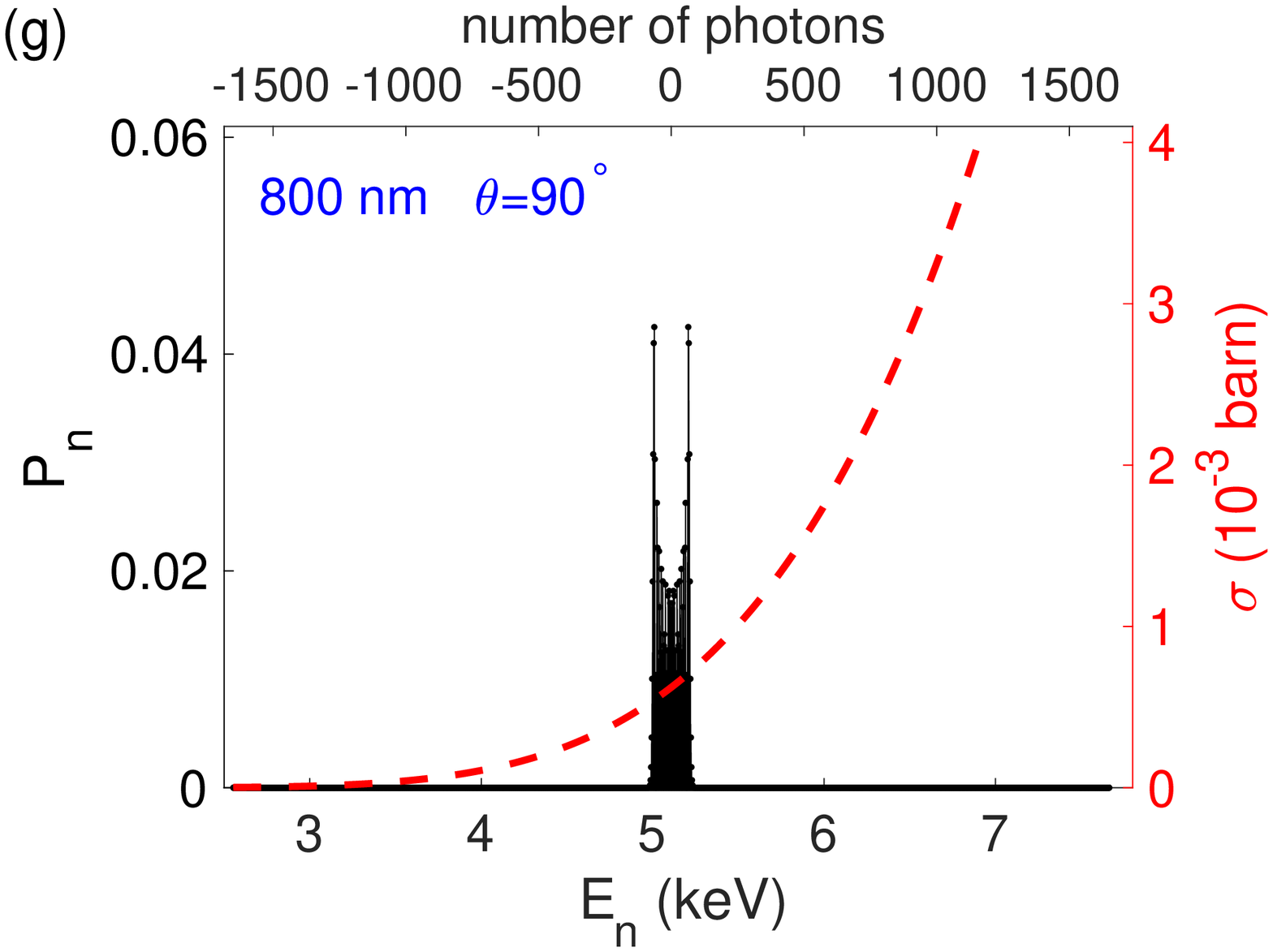}
 \hspace{0.1cm}
 \includegraphics[width=5cm,trim=0 0 0 0]{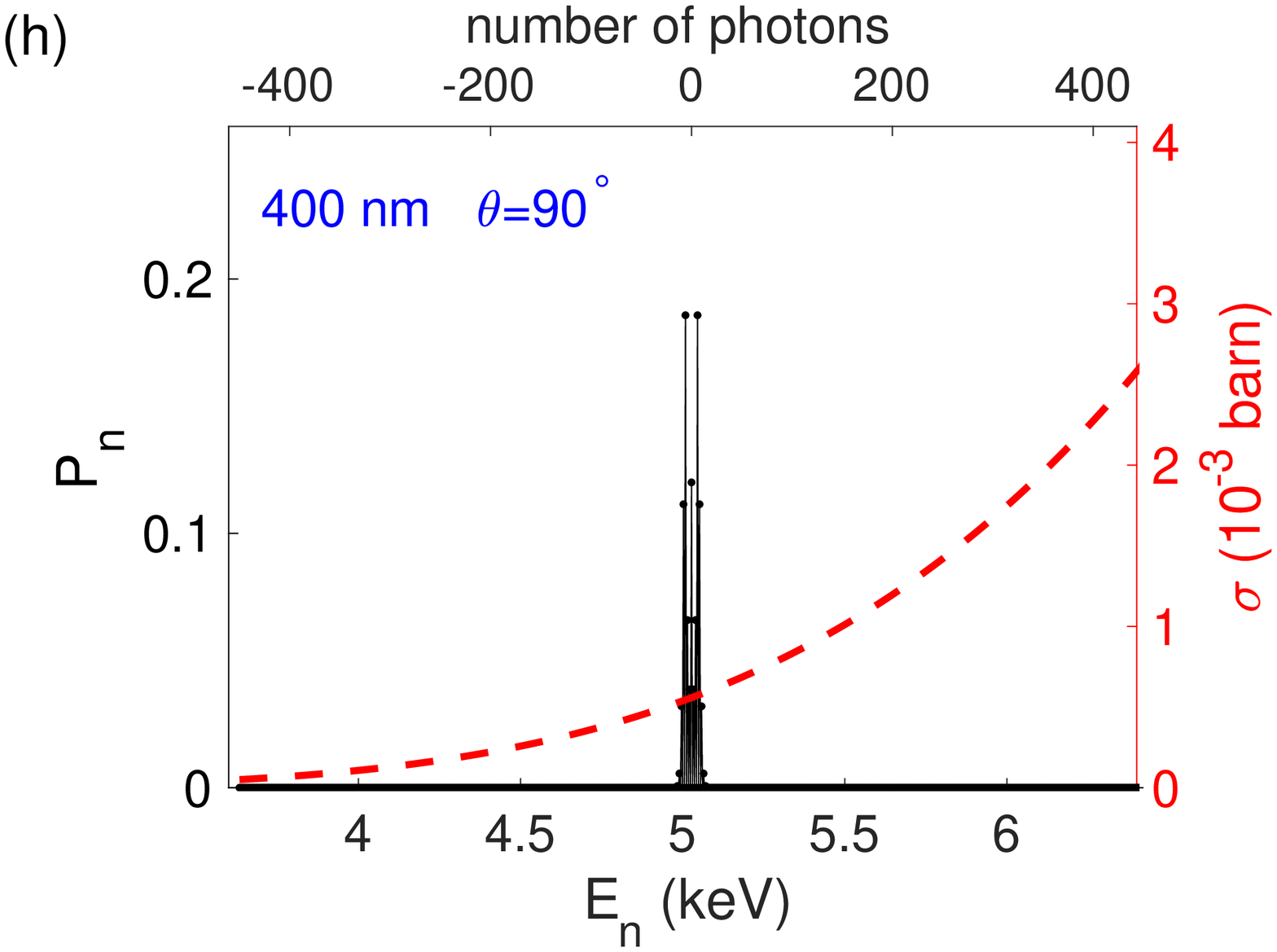}
 \hspace{0.1cm}
 \includegraphics[width=5cm,trim=0 0 0 0]{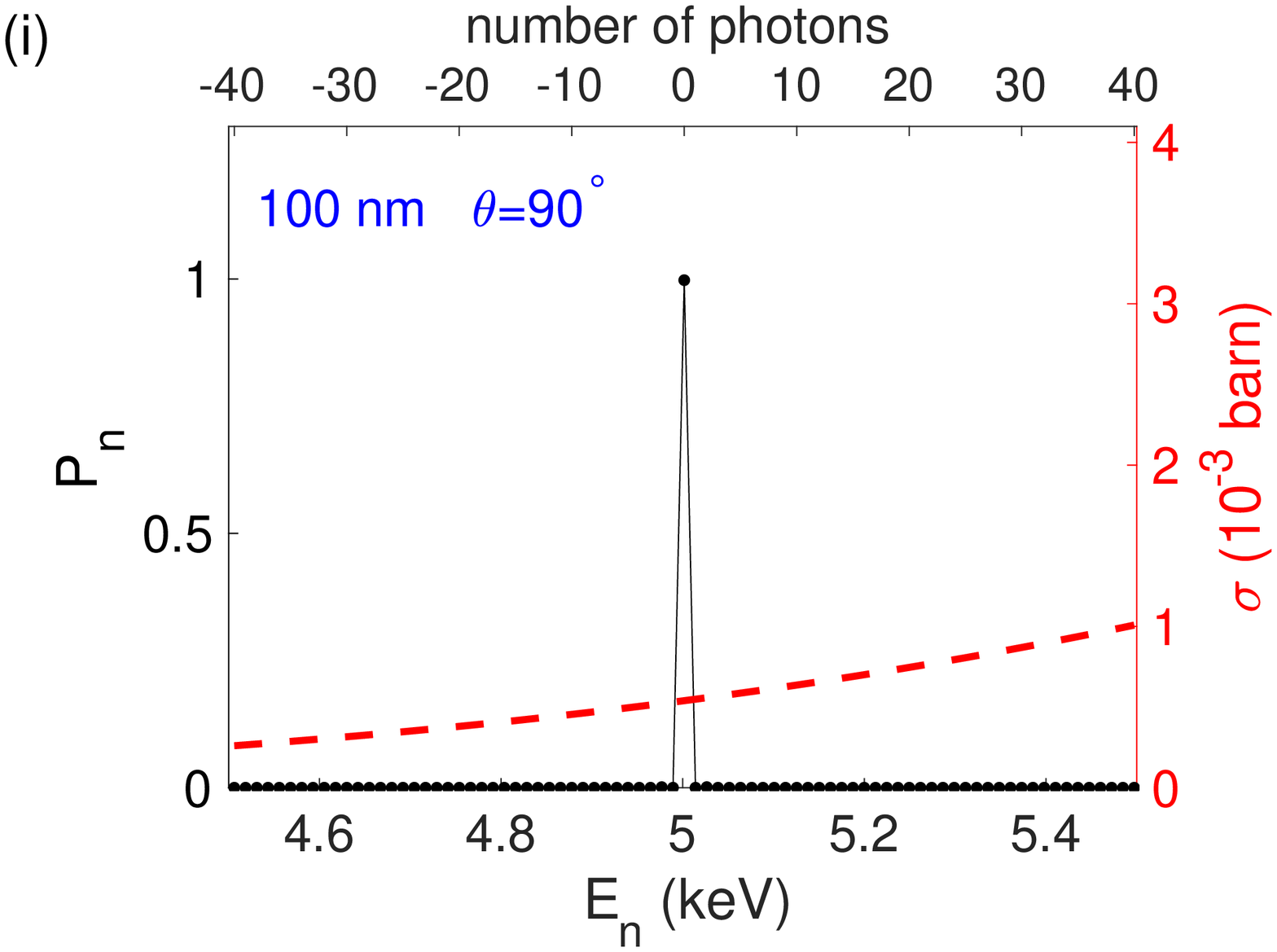} 
 \caption{Distributions of $P_n$ for laser wavelength 800 nm (left column), 400 nm (middle column), and 100 nm (right column). The laser intensity is $1\times10^{20}$ W/cm$^2$ for all the three wavelengths. For each wavelength three $\theta$ angles are shown, as labeled on figure. $\sum_{n=-\infty}^{\infty} P_n = 1$ holds for each figure. The fusion cross section $\sigma(E)$ has also been shown in each panel (red dashed curve) for the corresponding energy range, in linear scale (right axis).}\label{f.Pn}
\end{figure*}

{\it The Volkov state and its energy distributions.} In the center-of-mass frame, the two-body DT system is described by a single particle with mass $\mu = m_1 m_2/(m_1+m_2)$. Here we use subscript 1 for deuteron, and 2 for triton. This relative-motion particle with energy $E$ can be described asymptotically by a plane wave
\beq
\psi(\bm{r},t) = \exp\left\{ i\bm{p}\cdot \bm{r} - i Et \right\},
\eeq
where the momentum has magnitude $p=\sqrt{2\mu E}$. 

In the presence of a laser field, the asymptotic state of the particle becomes a Volkov state \cite{Volkov-35}
\beq
\psi_V(\bm{r},t) = \exp \left\{ i\bm{p}\cdot \bm{r} - i E t - i \int_0^t H_I(t') dt' \right\},  \label{e.psiV1}
\eeq
where $H_I$ is the interaction Hamiltonian with the laser field
\beq
H_I(t) = -\frac{q}{\mu} \bm{p} \cdot \bm{A}(t) + \frac{q^2}{2\mu} A^2(t). \label{e.Hi}
\eeq
Note that $q = (q_1 m_2 - q_2 m_1)/(m_1+m_2) = 0.2e$ is an effective charge for relative motion, and $\bm{A}(t) = \hat{z} A_0 \sin \omega t$ is the vector potential of the laser field, assumed to be linearly polarized along the $z$ axis. We neglect the spatial variation of the vector potential and further discussions on this point will be given later. Also because we are dealing with intense laser fields, the $A^2(t)$ term is kept. 

The Volkov wavefunction, with the above-given vector potential, can be expanded in terms of the photon numbers
 \beq
 \psi_V(\bm{r},t) = e^{ i\bm{p}\cdot \bm{r} } \sum_{n=-\infty}^{\infty} e^{iu} F_n(u,v) e^{  -i (E+U_p+n\omega) t }.  \label{e.psiV3}
 \eeq
And the coefficient $F_n(u,v)$ is calculated via
\beq
F_n(u,v) = \frac{1}{2\pi} \int_{-\pi}^{\pi} e^{ -i u \cos \xi + i v \sin 2\xi + i n\xi } d\xi.
\eeq
For convenience we have defined $U_p = q^2 A_0^2/4\mu$ which is the ponderomotive energy, $u = u(\theta) = qpA_0\cos\theta/\mu\omega$, and $v = q^2 A_0^2/8\mu\omega$. Here $\theta$ is the angle between $\bm{p}$ and the $+z$ axis, and $\theta$ enters into the formalism through $u$. In a thermal environment the direction between the particle momentum $\bm{p}$ and the laser polarization axis (the $z$ axis) is random.

In the laser field, the charge particle does not have a definite energy. Instead, it has a series of possible energies $E_n \equiv E + U_p + n\omega$ for all integers $n$. That is, the particle can absorb or emit integer numbers of photons in the laser field. The probability with energy $E_n$ is $P_n (u,v) = \left| F_n(u,v) \right|^2$, and the total probability is equal to unity $\sum_{n=-\infty}^{\infty} P_n = 1$.

Some example $P_n$ distributions are given in Fig. \ref{f.Pn}, for laser wavelengths 800, 400, and 100 nm. The same intensity of $1\times10^{20}$ W/cm$^2$ is used. The bare energy (energy without laser fields) $E = 5$ keV, corresponding to a temperature of about 55 million kelvin, a typical temperature in controlled fusion research. For each wavelength, three $\theta$ angles, namely, $0^\circ$, 45$^\circ$, and 90$^\circ$ are shown, as labeled on each figure.

The most important feature is that higher $E_n$ components are easier to be populated with longer wavelengths. For 800 nm and $\theta = 0^\circ$, $E_n$ with $-1400<n<1400$ are populated, as shown in Fig. \ref{f.Pn} (a). The photon energy for this wavelength is 1.55 eV, and the populated range of energy is 3.0 keV $<E_n<$ 7.3 keV (The ponderomotive energy $U_p$ is about 0.1 keV). For 400 nm, although the photon energy doubles, the populated number of photons is $-350<n<350$, and the populated range of energy is 4.0 keV $< E_n <$ 6.2 keV, as shown in Fig. \ref{f.Pn} (b). For 100 nm, only $-25 < n < 25$ or 4.7 keV $< E_n <$ 5.3 keV are populated as shown in Fig. \ref{f.Pn} (c). As $\theta$ increases from $0^\circ$ to $90^\circ$, the populated range of $E_n$ decreases, as can be seen by comparing each column of Fig. \ref{f.Pn}. $P_n$ for $\theta>90^\circ$ is the same as that for ($180^\circ-\theta$).

Another important feature is that $P_n$ does not decrease with increasing $|n|$. Instead, the general trend is that $P_n$ increases with $|n|$ (while fluctuating), reaches two high peaks near the maximally populated $|n|$, then terminates. This means that the particle can have substantial probabilities being with energies that are considerably higher (or lower) than its bare energy.  

In short, intense low-frequency laser fields are highly effective in transferring energy to the DT system. This will lead to substantially enhanced fusion probabilities, as will be shown below.

\begin{figure} [t!]
 \centering
 \includegraphics[width=6cm,trim=0 0 0 0]{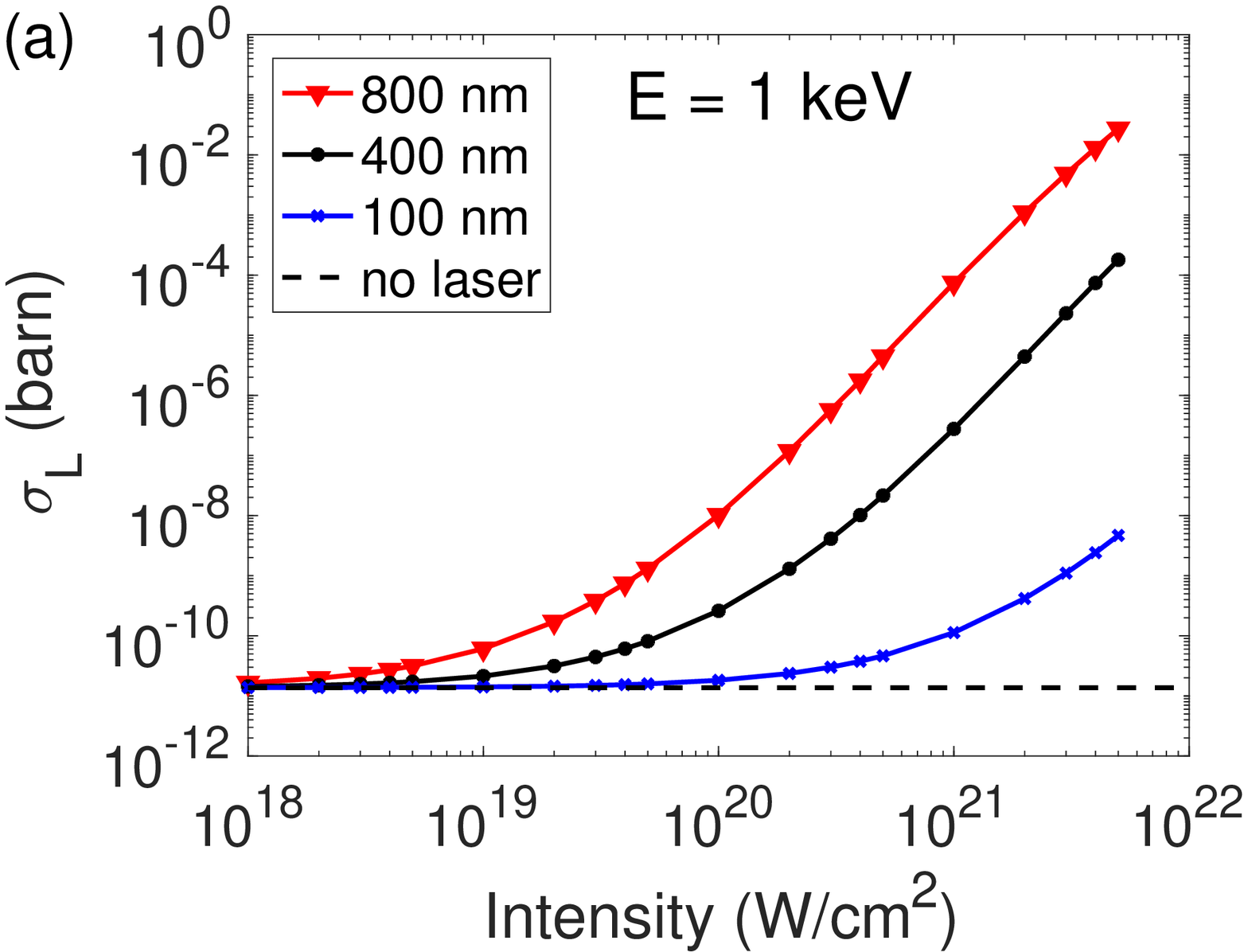} \\
 \includegraphics[width=6cm,trim=0 0 0 0]{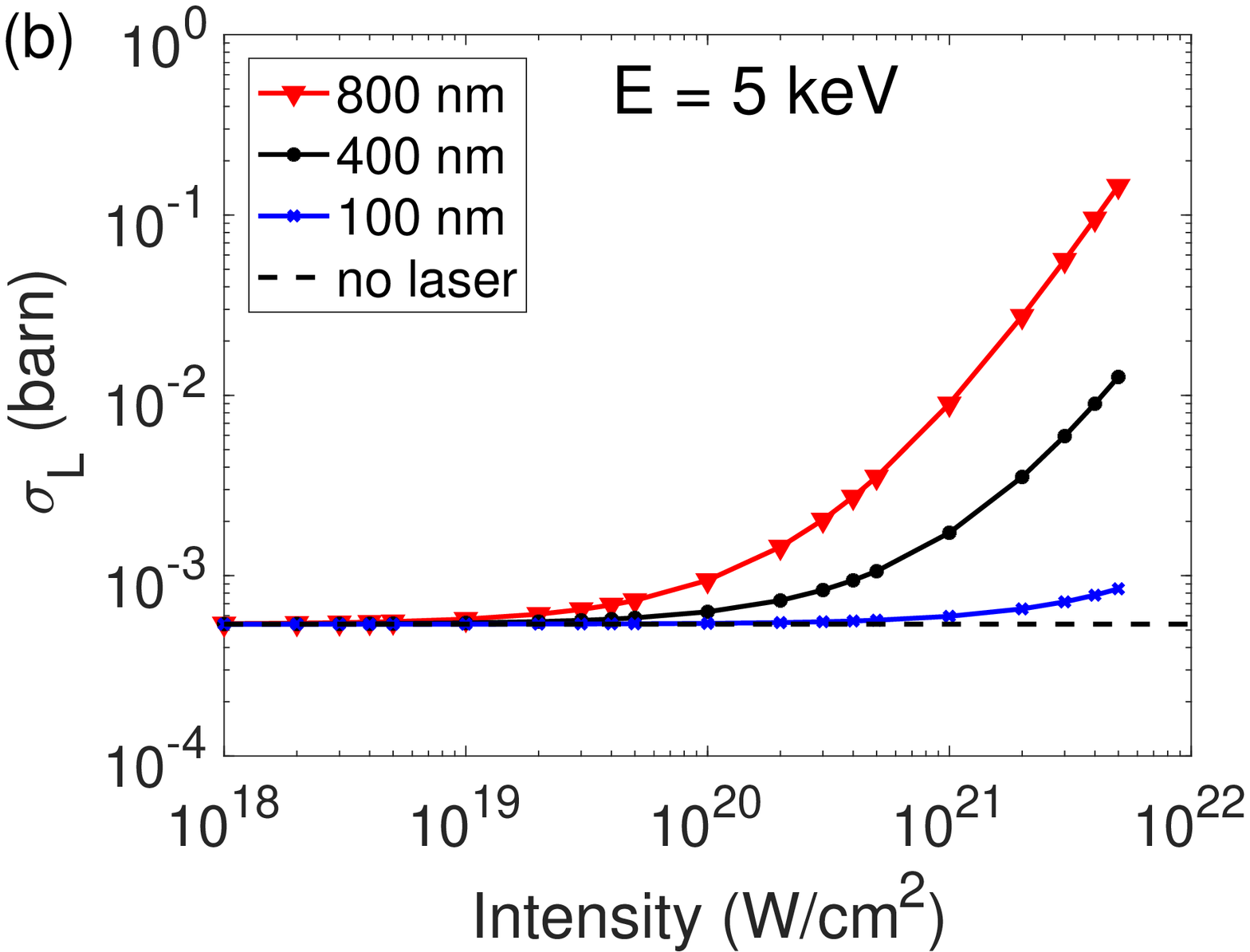} \\
 \includegraphics[width=6cm,trim=0 0 0 0]{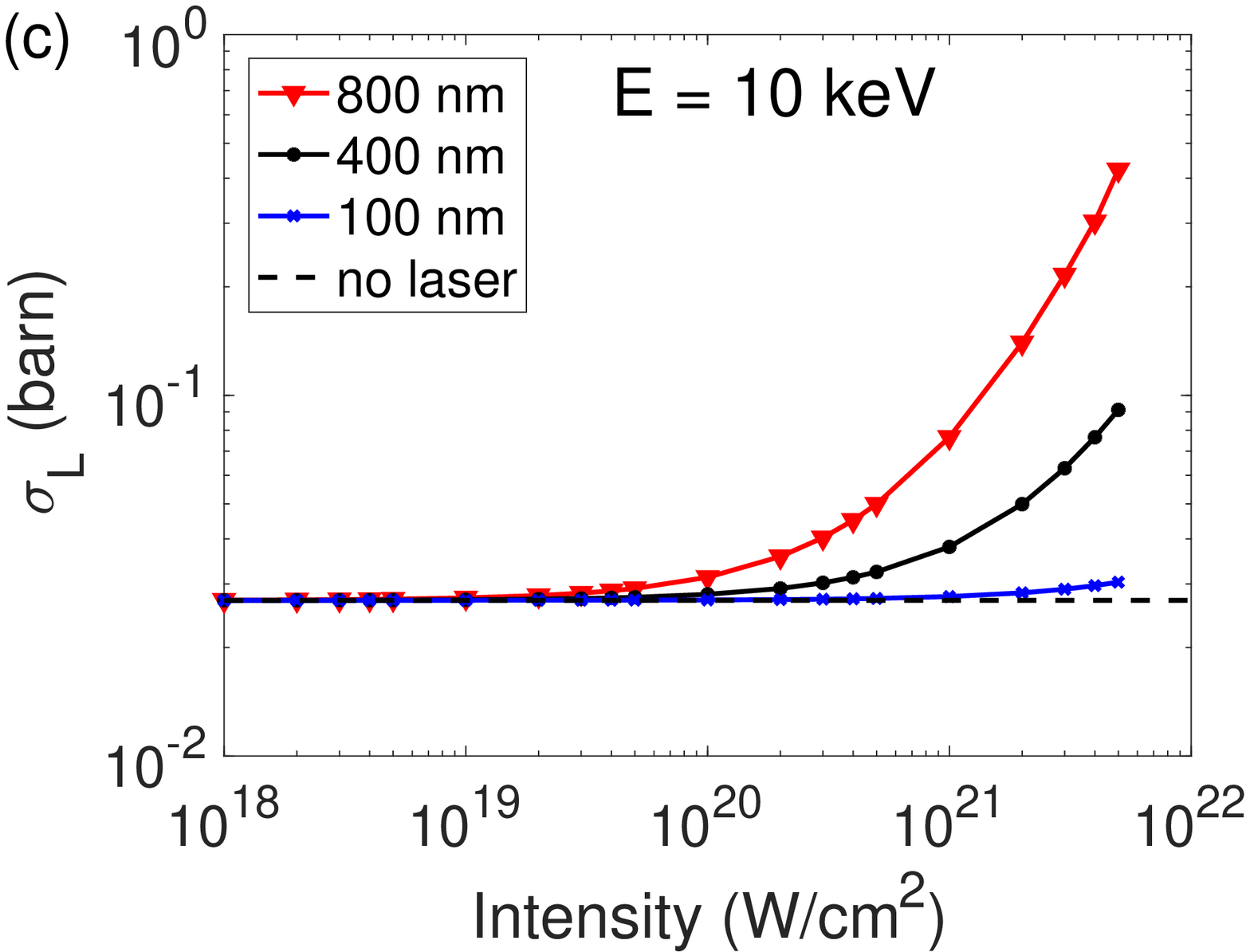} 
 \caption{Angle-averaged effective DT fusion cross section $\sigma_L$ under different laser intensities and wavelengths for $E=$ (a) 1 keV, (b) 5 keV, and (c) 10 keV. The horizontal dashed line in each figure marks the corresponding laser-free cross section.}\label{f.sigmaL}
\end{figure}

{\it Enhanced DT fusion probabilities in intense laser fields.} A component $E_n$ with $n>0$ will lead to a higher fusion probability than the bare energy $E$ does. The corresponding component $E_{-n}$ will lead to a lower fusion probability than the bare energy $E$ does. A net gain, however, can be obtained due to the exponential dependency of the DT fusion cross section on the energy, especially at relatively low energies, as can be seen from Fig. \ref{f.sigmaSE}(a) and also from Fig. \ref{f.Pn} in linear scale (red dashed line in each panel). This is the mechanism of enhanced DT fusion probabilities in intense laser fields.

To be more quantitative, given the laser parameters (intensity and wavelength) we may define an effective DT fusion cross section for each angle $\theta$ as
\beq
\sigma_L (E, \theta) =   \sum_{n=-\infty}^{\infty} P_n[u(\theta),v] \sigma(E+U_p+n\omega),
\eeq
and an angle-averaged effective fusion cross section as
\beq
\sigma_L(E) = \frac{1}{2} \int_0^\pi \sigma_L (E, \theta) \sin\theta d\theta.
\eeq
It is to be emphasized that the laser field does not change the cross section function $\sigma(E)$ {\it per se}, it just changes the energy of the particle before tunneling.

Fig. \ref{f.sigmaL} shows angle-averaged $\sigma_L$ under different laser intensities and wavelengths, for $E = 1$, 5, and 10 keV. These energies cover typical range of temperatures in controlled fusion research. One can see from all the three energies that $\sigma_L$ are substantially higher than the corresponding laser-free cross sections. The stronger the laser intensity, the larger the $\sigma_L$. For $E = 1$ keV, wavelength 800 nm, and intensity $5\times10^{21}$ W/cm$^2$, the enhancement is over 9 orders of magnitude. For $E = 5$ keV and the same laser parameters, the enhancement is over 2 orders of magnitude. For $E = 10$ keV and the same laser parameters, the enhancement is over 1 order of magnitude. The factor of enhancement drops as $E$ increases because the cross section function $\sigma(E)$ [Fig. \ref{f.sigmaSE}(a)] increases more slowly as $E$ increases.

One also sees from Fig. \ref{f.sigmaL} that longer wavelengths are more efficient in enhancing the fusion probability. The factor of enhancement drops as the wavelength changes from 800 nm to 400 nm to 100 nm. This is a direct consequence of the probability distribution $P_n$ explained above in Fig. \ref{f.Pn}: It is easier to populate high $E_n$ components using longer wavelengths.

Without laser fields, the DT fusion cross section for 1 keV ($1.37\times10^{-11}$ barn) is over 9 orders of magnitude smaller than that for 10 keV (0.027 barn). This gap can be filled, to a large extent, by intense laser fields. For example, the angle-averaged $\sigma_L$ with laser wavelength 800 nm and intensity $5\times10^{21}$ W/cm$^2$ is 0.0272 barn for 1 keV, and 0.423 barn for 10 keV. The gap shrinks to about 1 order of magnitude. Therefore it should be possible to relax the DT-fusion temperature requirement, which is known to be difficult to achieve and maintain, by using intense low-frequency laser fields.

{\it Further remarks.} In the above analyses the laser field does not alter the DT fusion process from the fundamental level, such as affecting the $S$ function. This is justified by the fact that what we now regard as very intense laser fields, such as those with intensities on the order of $10^{21}$ W/cm$^2$, are still negligible compared to nuclear potentials. The laser field has little effect on processes inside a nucleus or when the deuteron and triton are very close to each other. The role of the laser field is to change the particle energy before tunneling, and we have shown that this energy change has a substantial effect on the fusion probabilities.

For simplicity we do not include the spatial variation of the laser field. This approximation holds when the spatial range of motion of the particle is much smaller than the spatial range across which the laser field amplitude changes appreciably. The former range can be estimated by the quiver motion amplitude of the particle $qA_0/\mu\omega$, and the latter range can be estimated by the radius of the laser focal spot $R_c$. The required condition is $qA_0/\mu\omega \ll R_c$. For example, for $R_c\sim 1$ $\mu$m and intensity $5\times10^{21}$ W/cm$^2$, the frequency needs to satisfy $\omega \gg 0.0013$ a.u., or the wavelength $\lambda \ll 3.5\times10^4 $ nm. The wavelengths used in this paper obviously satisfy this condition. 

We also mention that the Volkov state has been widely used and very successful in strong-field atomic physics, the research discipline studying the interaction between atoms and intense laser fields, to describe the state of an ionized electron \cite{Keldysh-65, Faisal-73, Reiss-80, Lewenstein-94}.

Extension to elliptically or circularly polarized laser fields is straightforward. We find that elliptical or circular polarization does not lead to higher efficiencies in enhancing the DT fusion probabilities, mainly due to the reduction of the laser amplitude from $A_0$ to $A_0/\sqrt{1+\varepsilon^2}$ with $\varepsilon$ the degree of ellipticity. 

{\it Conclusion.} In conclusion we have considered the physics of using intense laser fields to enhance the DT fusion probabilities. We have answered the questions whether and by how much the DT fusion probabilities can be enhanced by intense laser fields, especially those with frequencies in the near-infrared regime for the majority of intense laser facilities. The combination of high intensity and low frequency leads to highly effective energy transfer from the laser field to the DT system, due to energy properties of the quantum Volkov state. The results show that the probabilities of DT fusion can be substantially enhanced, by at least an order of magnitude, in 800-nm lasers with intensities on the order of 10$^{21}$ W/cm$^2$. The results also show that low-frequency lasers are more efficient in enhancing DT fusion than high-frequency lasers. Our results indicate that intense low-frequency laser fields can be very helpful to controlled fusion research, and the demanding temperature requirement may be relaxed if intense low-frequency laser fields are fully exploited.

{\it Acknowledgements.} The author thanks Prof. J. H. Eberly for reading the manuscript and providing helpful suggestions. This work was supported by Science Challenge Project of China No. TZ2018005, NSFC No. 11774323, and NSAF No. U1930403.

\end{document}